\documentclass[journal,comsoc]{IEEEtran}
\usepackage[T1]{fontenc}
\usepackage{multirow}

\ifCLASSINFOpdf
\else
\fi
\usepackage{amsmath}
\usepackage[cmintegrals]{newtxmath}
\usepackage{url}

\usepackage{graphicx}

\begin{document}
%
\title{Automated Big Traffic Analytics for Cyber Security}
%
%
%

\author{Yuantian~Miao,
Zichan~Ruan, 
Lei~Pan,~\IEEEmembership{Member,~IEEE,} 
Yu~Wang,~\IEEEmembership{Member,~IEEE,} 
Jun~Zhang,~\IEEEmembership{Member,~IEEE,} 
and~Yang~Xiang,~\IEEEmembership{Senior Member,~IEEE}
\thanks{Y.~Miao, J.~Zhang, and Y.~Xiang are with the School of Software and Electrical Engineering, Swinburne University of Technology, John Street, Hawthorn, Victoria 3122, Australia. E-mail: (\{ymiao, junzhang, yxiang\}@swin.edu.au)

Z.~Ruan and L.~Pan are with the School of Information Technology, Deakin University, Geelong, VIC 3220, Australia. E-mail: (\{zichanr, l.pan\}@deakin.edu.au)

Y.~Wang is with School of Computer Science, Guangzhou University, 230 Guangzhou University City Outer Ring Road, Guangzhou 510006, China. Email: yuwang@gzhu.edu.cn
}}

\maketitle

\begin{abstract}
Network traffic analytics technology is a cornerstone for cyber security systems. We demonstrate its use through three popular and contemporary cyber security applications in intrusion detection, malware analysis and botnet detection. However, automated traffic analytics faces the challenges raised by big traffic data. In terms of big data's three characteristics --- volume, variety and velocity, we review three state of the art techniques to mitigate the key challenges including real-time traffic classification, unknown traffic classification, and efficiency of classifiers. The new techniques using statistical features, unknown discovery and correlation analytics show promising potentials to deal with big traffic data. Readers are encouraged to devote to improving the performance and practicability of automatic traffic analytic in cyber security.
\end{abstract}


%
\IEEEpeerreviewmaketitle

\section{Introduction}

\IEEEPARstart{A}{s} an increasingly huge amount of important information about users are delivered and stored on the Internet, cyber security becomes a primary concern.  Protecting network from various attacks is the paramount task, where traffic data analysis is a key technology \cite{nguyen2008survey}. Moreover, traffic classification applying data analytics significantly improves the effectiveness and efficiency of the process, as well as enables us to detect abnormal traffic patterns. 

\begin{figure}[!t]
\center
  \includegraphics[scale=0.5]{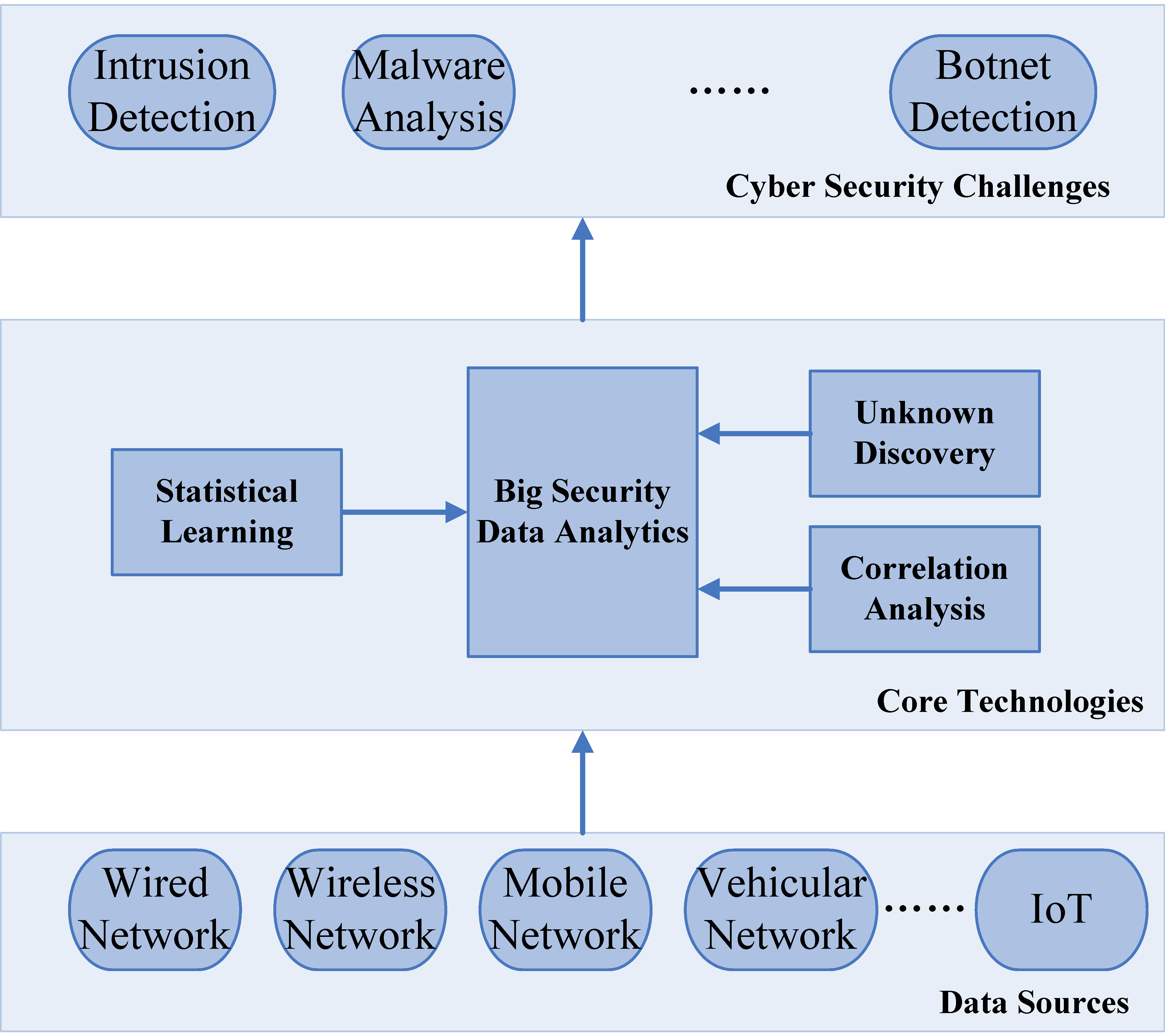}
\caption{Traffic analytics for cyber security}
\label{fig:f1structure}       
\end{figure}

Some popular applications of cyber security involved with traffic data analysis include intrusion detection, malware analysis and botnet detection. The automatic techniques based on machine learning algorithms are proposed for these applications. To be specific, intrusion detection system can catch the abnormal traffic including peer-to-peer (P2P) malicious traffic, denial-of-service (DoS) attack traffic and spams;  malware can steal user information and cause harms to users. In addition, mobile malware spread widely \cite{gui2016analysis}. To detect such traffics, malware behaviors are analyzed and this knowledge promotes traffic classification in identifying the corresponding malware \cite{gui2016analysis}; botnets consist of many Internet hosts controlled and manipulated by botmakers via the infection of malware \cite{dainotti2012analysis}. As for this bot detection, the list of command and control (C\&C) domain is used as a reference among bots identification. As a fundamental technique among these attack detections, traffic classification is applied widely in recognizing various network traffic and singling out the abnormal traffic.

Considering the massive data existed in network traffic, the primary problem of traffic classification is raised by big data which can be illustrated with three characteristics --- volume, variety, and velocity \cite{suthaharan2014big, gui2016analysis}. 
Traffic classification is facing three key challenges, which respectively corresponds to real-time classification, unknown traffic classification, and the efficiency of automatic classification. 
In terms of real-time classification, statistical-based classification is more suitable than the port-based and payload-based approaches. 
Statistical-based classification considers the flow traffic as the unit object in analysis, whilst the other two methods build the classifier analyzing the packet information \cite{nguyen2008survey}. 
As for the unknown traffic problem, both supervised learning and unsupervised learning cannot solve the problem alone  \cite{zhang2015robust}. 
Therefore, Zhang et al.~\cite{zhang2015robust} proposed a robust network traffic classification with the combination of unsupervised learning and supervised learning to resolve this problem. 
Lastly correlation technique is often used to improve the classification performance \cite{ma2006unexpected, wang2014internet}. In summary, the core concepts mentioned above is depicted in Figure~\ref{fig:f1structure} with the relationship among data sources, core technologies and cyber security challenges.

\section{Cyber Security Applications with Traffic Analytics}
Most cyber security systems rely on the deep understanding of network traffic characteristics. 
In this section, we illustrate three typical applications of cyber security involving traffic analytics including intrusion detection, malware analysis, and botnet detection.

\subsection{Intrusion detection}
An intrusion detection system (IDS) aims at recognizing malicious traffic from normal traffic. To achieve this goal, IDS  scans current traffic before rerouting. Because of the heterogeneous nature of the attacks, the malicious traffic is often embedded in  botnet traffic, DoS attack traffic, spam traffic, and so on. The accuracy of detecting such malicious traffic is often low. In general, the target information of inspecting traffic includes IP addresses, ports, and payload data. By analyzing the information, abnormal traffic data can be discovered by an IDS. Then traffic classification is applied to separate these abnormal traffic among existing traffic. Accordingly, alerts will be sent by the IDS. 

We use an example to show how traffic analytics is used in intrusion detection. Ling et al.~\cite{ling2015torward} developed \emph{TorWard} as an intrusion detection system for Tor.  Tor is the overlay network often used by attackers deploying botnet C\&C servers and/or sending spams because Tor encrypts the traffic and protects the attacker's privacy.  Specifically, Tor enforces source routing by choosing several Tor routers to establish an anonymous route along with the selected Tor routers. There is an exit router where all the routes pass through in the end. This exit router can be considered as a ``proxy" and contacts with the destination directly. In order to protect the network system through blocking possible harmful traffic, IP address and opened port can be configured in Tor manually according to specified policies. Unfortunately, It is a hard job for most Tor router administrators to know everything and everyone in the Tor network. Therefore, \emph{TorWard} is proposed by Ling et al.~\cite{ling2015torward} as an automated detection system in processing Tor intrusion discovery, classification and reaction.  The outbound traffic of Tor can be captured at the exit router. Before the traffic rerouted into Tor, an IDS is positioned on the NAT gateway.  Ling et al.~revised the Tor source code and maintain rules of firewall dynamically in order to except the hinder from non-Tor traffic.

Additionally, Ling et al.~\cite{ling2015torward} found that nearly 10\% Tor traffic are alerted by IDS because of its malicious traffic generated by botnet, DoS attack traffic, spams, and so on. Moreover, they also designed a defense function to limit the usage of Tor by blocking intrusion traffic. Such an IDS checks the suspected source IP and ports, and if this traffic instance is classified as intrusion, then the tear-down command is expected to be sent to the Tor exit routers. And with extensive rule setting, IDS can block numerous intrusion traffic. Furthermore, a signaling-based method named dual-tone multi-frequency is suggested by Ling et al.~\cite{ling2015torward} to search the correlated botnet traffic from exit routers. They analyzed the intrusion traffic in Tor and summarized a category with possible IP of C\&C servers, well-known malware traffic, doubtful DNS query, spams and suspicious IRC traffic. Specific traffic classification methods are used on two types of signature-based IDS --- Suricata and Snort \cite{ling2015torward}. Suricata is configured to store alert messages in binary format, and Snort employs Unix domain sockets.  Tor traffic can be generally grouped as two traffic types: inbound and outbound. 
Because the characteristic of inbound traffic sets captured by IDS transmitting between endpoints, traffic classification technique equipped with classifying encrypted data is needed. With a powerful traffic classification, IDS can be effective and reliable for automatically detecting the potential malicious traffic.

\subsection{Malware analysis}
The number and the level of technological sophistication of malicious software are increasing remarkably, as a consequence of the outstanding involvement of smartphone technologies. Thus, malware analysis is an important task in cyber security. Alerts reported  by malware detection system related to malicious traffic include \emph{unclassified}, \emph{misc-attack}, \emph{Trojan-activity}, \emph{not-suspicious}, and \emph{misc-activity} \cite{ling2015torward}.  A well-known example of mobile malware in iOS is \emph{XcodeGhost}, which was reported as the reason of numeric user privacy-leakage in the late 2015,  drew the attention of the society and aware people the importance of cyber security \cite{gui2016analysis}. Xcode is a development toolkit for iOS applications. However, \emph{XcodeGhost}, a malicious version of Xcode, was uploaded to a Chinese shared cloud service \emph{Baidu cloud}. Application developers downloaded the \emph{XcodeGhost} without being aware of its danger, developed the infected applications and published them in Apple's App Store. Anyone who downloads and starts those infected applications in their devices could be a victim to the privacy-leakage.

Network traffic analytics is an important tool to identify infected applications in a large scale.  Let us see how traffic analysis is used to study the behavior of \emph{XcodeGost}. There are two mainstream methods to study the threat of \emph{XcodeGhost}, the first method is from the view of source code of \emph{XcodeGhost} to find the detailed information of the malware mechanism, the other is scanning the source code of iPhone Application (IPA) packages to detect the infected applications. However, both methods fail to provide sufficient information about the \emph{XcodeGhost} such as the number and ratio of infected devices, the time and network traffic flow volume characteristics of \emph{XcodeGhost}-related HTTP requests and the infected applications \cite{gui2016analysis}. Combining the fingerprint of applications running online and its web-knowledge, a novel method was built up to identify the infected applications instead of scanning the source code of IPA. This mechanism consists of 5 steps --- 1) extract fingerprints from those applications, 2) collect web-knowledge by analyzing extracted information, 3) merge data and identify applications, 4) check those applications manually, and 5) compare with the identification results. Massive network traffic were analyzed to gain flow statistical characteristic of \emph{XcodeGhost}, which may be useful in grasping the operating mechanism of \emph{XcodeGhost} and understanding other malware like \emph{XcodeGhost}. To classify HTTP traffic with specific applications, the characteristics of HTTP headers are important for analysis, however, the extracted fingerprint information is not always clear or complete. The unique identifier of an application associated with advertisement traffic and integrated with multiple HTTP requests is needed to extract the fingerprint information and identify the full details of an application.

To fully investigate \emph{XcodeGhost}, Gui et al.~\cite{gui2016analysis} proposed a novel method by examining the traffic statistical characteristics of applications. The researchers explored a large volume of real-world network traffic and found that 930 million out of 1,550 million iPhone devices were infected within 232 days, which means more than 60\% of devices potentially leaked the privacy information. The results were critical: 842 applications were identified as \emph{XCodeGhost} infected by the new proposed method in traffic network, some of them are famous and popular applications throughout China like \emph{Wechat}, \emph{railway 12306}, \emph{didi taxi}, \emph{carrot fantasy} and so on \cite{gui2016analysis}. The results suggested that those infected applications  send privacy information of users to a unique web server. Such behavior could not be detected by using traditional detection methods because the privacy leaking messages are hidden inside many legitimate HTTP requests.

\subsection{Botnet detection}
Botnets enable attackers to send spams, launch distributed denial-of-service (DDoS), run brute-force password cracking, steal private information, and hide the origin of cyber attacks \cite{dainotti2012analysis}. Moreover, the malware traffic can be spread rapidly through this platform. Hence, botnet detection is an important part in cyber security. According to the structure of botnets, there are two  categories ---- P2P botnet and centralized botnet \cite{ling2015torward}. In a P2P botnet, the botmaster can control each bot with distributed commands sent from peers; in a centralized botnet, the centralized C\&C architecture is formed with protocols like IRC and HTTP. Comparing to the centralized botnet, P2P botnet is relatively feasible regardless of its complicated structure and costly management. From the point of view of user experience, centralized botnet is in extensive usage due to its simplistic structure, availability of source codes, and reusable codes.

Network traffic analytics plays an important role in the discovery and detection of botnets. The typical approach to detect bots and filter botnet traffic is to maintain a blacklist of discovered C\&C domains. However, the efficiency is poor because the blacklist is manually updated. Moreover, experienced botmasters often use robust P2P-based C\&C structures with domain generation algorithms to evade the detection by blacklisting and to increase the reliability of the botnet. That is, the bots search for working C\&C servers by periodically generating a set of pseudo-random domain names  and resolving the generated domain names to IP addresses through DNS queries. Therefore, these botnets can still survive even after some C\&C servers are detected and blocked.

Antonakakis et al.~\cite{antonakakis2012throw} show that the random domains generated by botmasters can be detected by analyzing DNS traffic. The key idea is that most random domains generated by domain generation algorithms and queried by the bots would receive Non-Existent Domain responses. Moreover, since the bots in the same campaign are using identical domain generation methods, they are very likely generating the same set of failed DNS queries. To take advantage of this characteristic, Antonakakis et al.~propose a traffic analytics based technique that combines supervised and unsupervised learning algorithms. The unsupervised algorithm is applied firstly to separate these random domains into several clusters according to the similarity of their make-ups of domain names. Then, supervised learning takes the input of the whole data set together with some labels of known domains. At last, a new unknown domain of bot can be detected if no label is assigned to a cluster.

\section{Key Challenges of Big Traffic Analytics}

\begin{figure}[!t]
\center
  \includegraphics[scale=0.5]{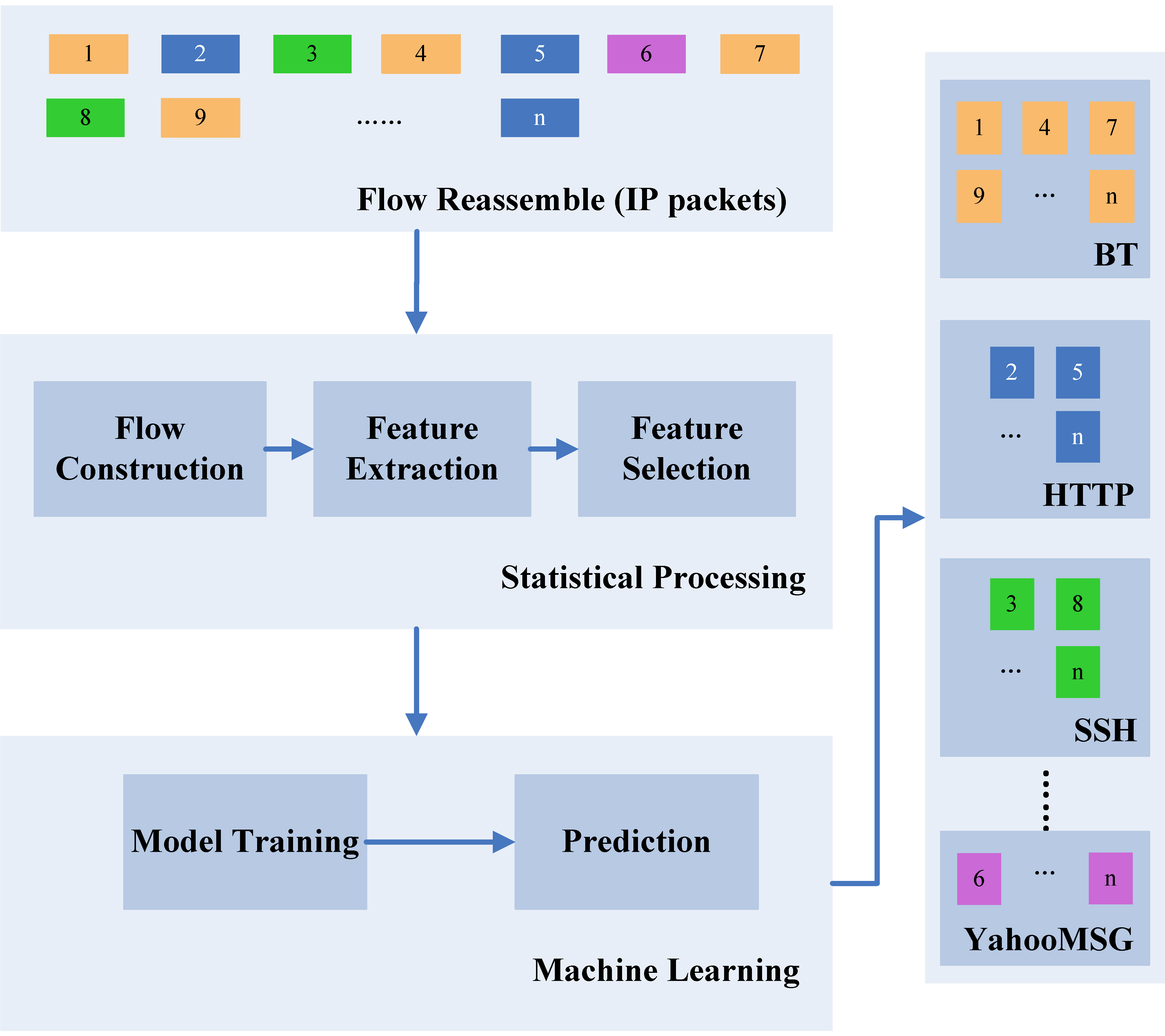}
\caption{The statistical-based analytics mechanism is presented in
  this figure. Flows are used as input of machine learning classifiers, after
  statistical processing is applied to the collected packets. The output shows
  the categorization results predicted by machine learning algorithms.} 
\label{fig:f2statistical}       
\end{figure}

As the previous section revealed, traffic analytics is a primary foundation for dealing with emerging issues in cyber security. This section lists the key components of traffic analytics and the associated challenges in the context of big data. We elaborate traffic classification in three key facets --- real time, robustness and efficiency corresponding to the three characteristics of big data featured in terms of volume, variety and velocity \cite{suthaharan2014big}.  

\subsection{Real time classification by using statistical features}

Statistical feature based traffic classification is as useful as port-based and payload-based classifications. Because the network traffic data grow
exponentially with the rapid deployment of high bandwidth
demanding services, the port-based and payload-based classifications
failed to meet the big data requirements
\cite{moore2005internet,nguyen2008survey,dahmouni2007markovian}.  
Specifically, the port-based classification identifies the
applications based on the port number extracted from 
the packets according to the IANA standards \cite{nguyen2008survey}. 
Because the port number can be easily modified or fabricated, the accuracy
of the port-based classification is too low to fulfill the requirements
of cyber security applications. Conversely, the payload-based
classification recognizes the protocol signature by inspecting the
payload of packets. Comparing with the port-based method, the
payload-based method produces more accurate results but at a
significantly higher computational cost. The statistical-based
classification balances accuracy and cost by analyzing and extracting
statistical characteristics from trafffic flows generated by
applications without inspecting the contents of individual
packets. Because of this holistic view, the statistical-based
classification can be used to analyze encrypted traffic widely used in
Virtual Private Networks, WiFi, Tor networks and so on.

Machine learning techniques are vital for the statistical-based
traffic classification. The traffic can be processed by supervised
learning, also known as classification, or by unsupervised learning,
also known as clustering.  Figure \ref{fig:f2statistical} depicts the
work flow --- the collected packets are firstly grouped into traffic
flows before certain statistical features such as packet size and
arrival time can be obtained and later selected; the data in the
feature space are then fed to the machine learning algorithm. Due to
the nature of network applications, the training data fed to the machine
learning algorithm can be inbalanced and/or polluted. An inbalanced data
set often consists of non-uniformly distributed packets of different
kinds; and a polluted data set often contains multiple wrongly labeled samples.    
The labeled data are used to train the machine learning algorithms,
after a certain amount of time, predictive models will be built up and
used as classifiers or predictors to classify testing data.

Generally, network traffic contains statistical features such as
packet size, inter-arrival time, flow idle time, distribution of
duration time, and so on, according to Nguyen and Armitage
\cite{nguyen2008survey}. These features can be used to successfully differentiate 
many Internet applications. If these flow statistical properties are
constructed as inputs, many machine learning algorithms can handle
the classification of real time traffic flow directly out of the box. For example, naive Bayes and decision
tree can build up a model for classification and prediction. Moreover, 
$k$-Nearest Neighbor algorithm is another statistical-based
classification method.

Nevertheless, the statistical-based classification is not
perfect. There are primarily three limitations. Firstly, it is
difficult and expensive to obtain a balanced training data set with the uniformly
distributed traffic classes from real-world networks. With an
unbalanced data set, the classes with a dominantly large number of instances are
trained more accurately than the rest. The implication can be
problematic when we try to identify a small portion of malicious
packets from a huge number of normal packets. According to \cite{biggio2012poisoning}, such adversary information 
decreases the effectiveness of machine learning remarkably. Secondly,
the statistical data that we collected and processed as 
training set might not be representative during the deployment period. For instance, data collected from 
one network may not be applicable else where. Thirdly, the presence of unknown
classes also affects the machine learning results. An unknown class is
a traffic class that existed in the testing data set but is not
included or recognized in the training data set. Therefore, there could be
error in classification as if a new class data occurs in the testing set.


\subsection{Robust classification by recognizing unknown classes}
Despite of the challenges in real time traffic classification, exploitation of zero-day vulnerabilities greatly affects the accuracy of traffic classification  \cite{zhang2015robust}. The reason is that statistical-based traffic classification relies on machine learning algorithms.  These classifiers cannot classify an unknown traffic which is caused by the exploitation of the zero-day vulnerabilities.  To recognize such attacks, our previous work \cite{zhang2015robust} proposed a Robust network Traffic Classification (RTC) in dealing with unknown classes that are absent from training data by integrating supervised and unsupervised learning methods. There are three processes in constructing RTC including unknown discovery, classification with unknown class, and system update. 

\begin{figure}[!t]
\center
  \includegraphics[scale=0.5]{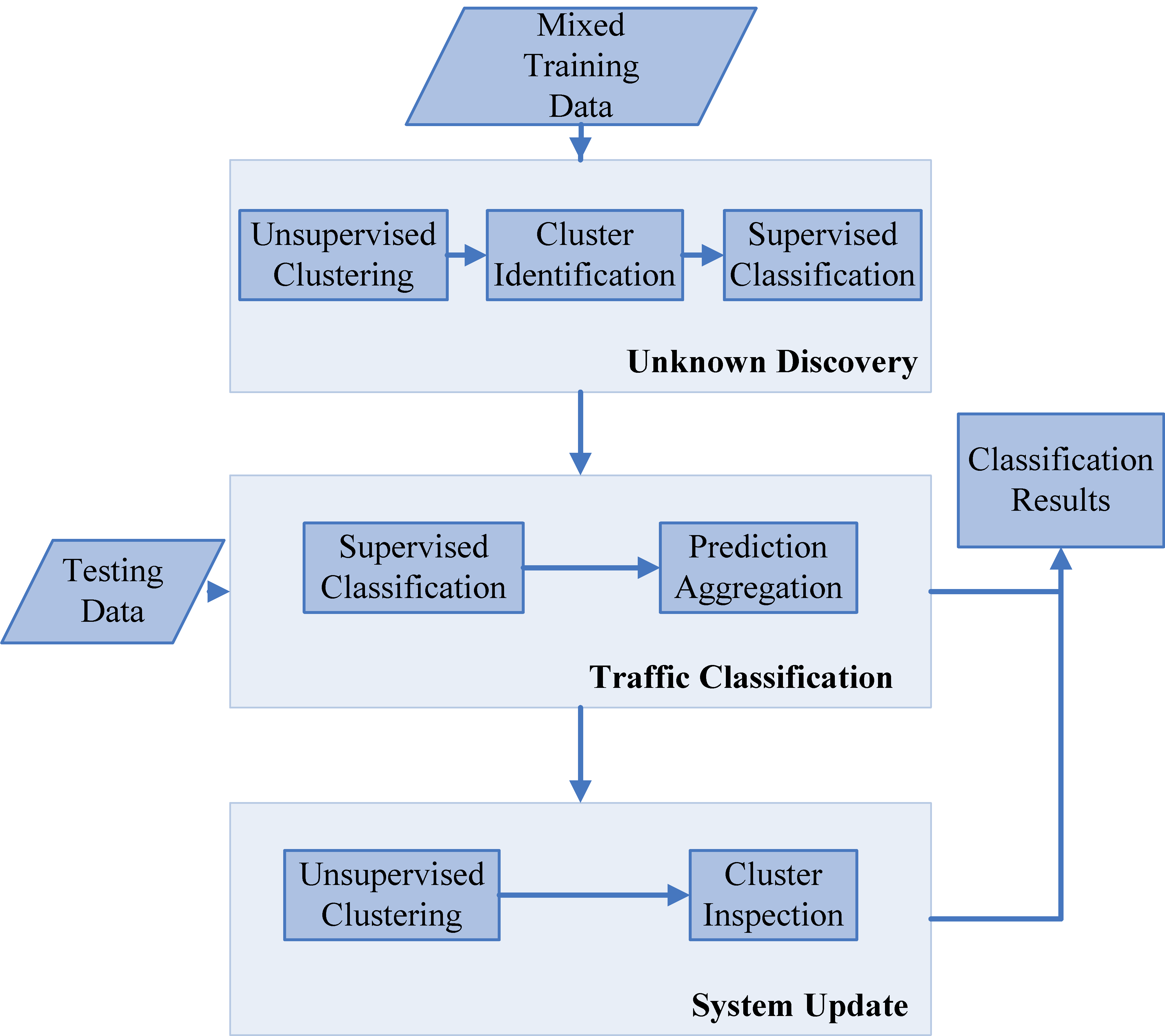}
\caption{The figure reveals the methods towards the challenge of unknown attacks when processing classification. Mixed training data is used as input of unknown discovery, which is an initial stage to label known classes, consequently, the unknown classes are outcropped. Predict the testing data with classifiers that derived from the clusters, and apply system update stage if necessary.}
\label{fig:f3unknown}       
\end{figure}

RTC has powerful classification ability in handling zero-day traffic's discrimination together with the traffic that the classifiers are familiar with.  Figure \ref{fig:f3unknown} shows the mechanism of RTC. In the first stage of RTC, unknown traffic are detected automatically prior to classification. The unlabeled samples are collected from a specific network obtaining the mixture of predefined traffic and new traffic. Then a two-step method is employed in distinguishing zero-day traffic from others. That is, an unsupervised learning method, $k$-means, is used to separate the mixture data into several clusters, while the labeled training instances help to identify the unknown traffic groups. And the zero-day class is captured in the unknown clusters with no or very few labeled instances assigned. Additionally, there may be more than one unknown application, thus the second step is to group them into a temporary super set for all unknown classes. After the processed  data are all labeled, RTC starts the classification stage by applying supervised learning algorithms with some heuristics to aggregate the prediction results.

In order to build an intelligent system with classifying class in fine-grained level, the traffic classification system should be updated when necessary. Specifically, the classifier should be equipped with the knowledge of unknown classes detected in the first stage instead of putting all unfamiliar traffic into a general category. In this third stage, $k$-means is applied again to cluster the previous data set. Then, several flows of each cluster are selected and inspected manually. If all the samples in one cluster are detected as unknown traffic, all flows in this cluster are acted as training inputs for classifier and a new class is labeled. After all new classes are found, they will be appended to previous known class set. Thus, the training data set is updated, whilst classifiers keep learning more unknown classes. 

Though RTC can recognize unknown classes, there are still rooms for improvement. In terms of zero-day attacks, flow correlation technique can be used together with RTC to increase the accuracy rate. One of the assumptions of RTC is that the distribution of traffic classes is static, which is not always true in the real-world networks. In addition, it is also a problem to accurately and precisely define an unknown class, especially when there are numerous new classes. Moreover, the cost of classification significantly increases with the growing size of training data set, whereas the accuracy rate will decrease rapidly with an increasing number of classes.


\begin{figure}[!t]
\center
  \includegraphics[scale=0.5]{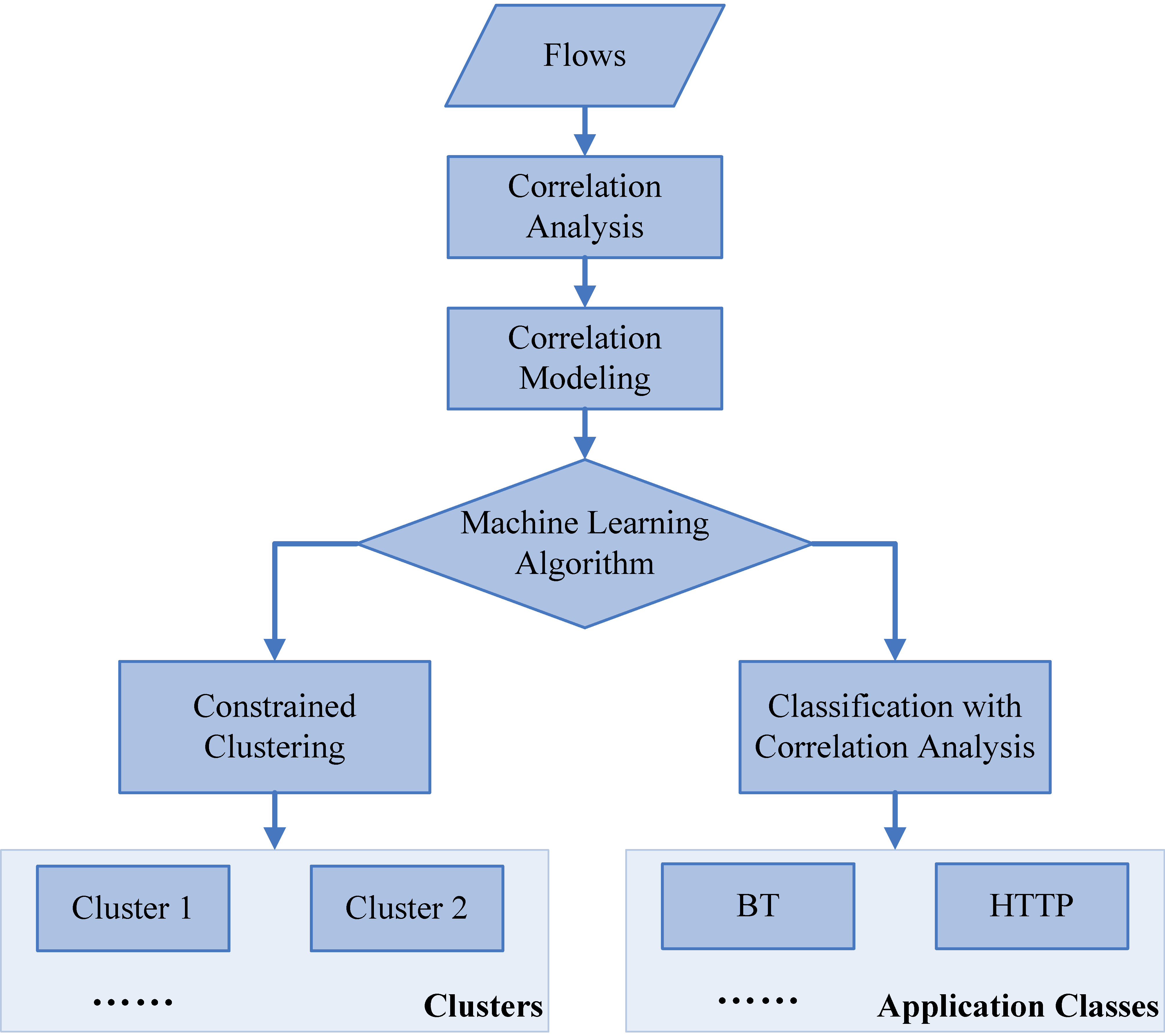}
\caption{The figure presents the mechanism of correlation analysis. Correlation models are proposed and merged in the machine learning algorithm after the correlation analysis of the flows. In the unsupervised clustering, the correlation model is applied as constraints when clustering and helps to improve the purity of each cluster. As for the supervised classification, application classes is better predicted after the correlation analysis combined with classification results.}
\label{fig:f4correlation}       
\end{figure}

\begin{table*}[!t]
\renewcommand{\arraystretch}{1.3}
\caption{Comparison of paper works related to above three parts}
\label{tab:compare}
\centering
\begin{tabular}{l l l l l }
  \hline
\multirow{ 2}{*}{Work} & Real-Time & Robustness & Correlation & Traffic Classifier \\
  & (volume) & (variety) & (velocity) & (machine learning) \\  
  \hline\hline
  \multirow{ 2}{*}{Moore and Zuev  \cite{moore2005internet}} & \multirow{ 2}{*}{Flow statistics based} & \multirow{ 2}{*}{Do not consider} & \multirow{ 2}{*}{Do not consider} & Naive Bayes Classifier \\
  &  &  &  & (supervised learning) \\
  \hline
   & & Handle Zero-day traffic & Correlated information: & Random forest \\
       Zhang et al.~\cite{zhang2015robust}        &           Flow statistics based   & Unknown discovery       & 3-tuple heuristic & $K$-means clustering \\
                                                               &                                   & 	       & (applied in BoF) 	& (compound learning) \\
  \hline
   &  &  &  & Product distributions \\
 \multirow{ 2}{*}{Ma et al.~\cite{ma2006unexpected}} & \multirow{ 2}{*}{Payload based}  & \multirow{ 2}{*}{Do not consider} & Correlated Information: & Markov Processes \\
   & &  & packet content & Common Substring Graphs \\
   &  &  &  & (unsupervised learning) \\
  \hline
   &  &  & Correlated Information: & Equivalence set constraints applied \\
  Wang et al.~\cite{wang2014internet} & Flow statistics based & Do not consider & 3-tuple heuristic & with $K$-Means cluster \\
   &  &  &  & (semi-supervised learning) \\
  \hline
   &  &  & Correlated Information: & Three NN classifiers \\
 \multirow{ 2}{*}{ Zhang et al.~\cite{zhang2013network}} & \multirow{ 2}{*}{Flow statistics based} & \multirow{ 2}{*}{Do not consider} & 3-tuple heuristic & combined with BoF: \\
   &  &  & (applied in BoF) &  AVG-NN, MIN-NN \& MVT-NN \\
   &  &  &  & (semi-supervised learning) \\
  \hline
\end{tabular}
\end{table*}

\subsection{Efficient classification by using correlation}
Correlation technique is applied in traffic classification to improve the classification efficiency without incurring too much processing time.  Correlation technique can help us to identify the traffic with little or minimal prior knowledge. An example is provided by Ma et al.~\cite{ma2006unexpected} where  correlation technique combined with the protocol model derived during the traffic classification. In their work, the correlated information of several unlabeled traffic flows is the same protocol described in their packets according to the same distribution of one protocol. Since not all flows in one application share one protocol, their correlation is partially correlated. As a result, they successfully classified the network flows automatically with only protocols provided. Another utilization of correlation information is based on the similar user behaviors \cite{dainotti2012analysis}.

Correlation technique not only can classify traffic data but also can improve clustering results by increasing each cluster's purity. In Wang et al.~\cite{wang2014internet}, correlated information are presented as equivalence set constraints which can reveal the flows with same protocols in application-layer and sharing for.  As shown in Figure~\ref{fig:f4correlation}, they used the unlabeled statistical features extracted from the raw packets in unsupervised clustering with correlation analysis applied in. In other words, the flow correlation information was used as constraints when clustering those data and helped increasing the purity of each clusters. Wang et al.~named the integrated method as Set-Based Constrained $K$-Means. Herein, the equivalence set are allocated with 3-tuple heuristics among the five features described for each flow including source IP address, source port number, destination IP address, destination port number, and the protocol. As a result, the classifiers after cluster identification combined with correlation analysis perform significantly better than those ones without correlation applied in terms of accuracy and computation time.

From Zhang et al.~\cite{zhang2013network}, correlation method also makes positive effects on supervised classification. The specific method they applied is named as Bag of Flow (BoF), which means the correlated traffic flows from one application are generated together as a bag. Specifically, three-tuple heuristic is considered in a short period of time, that is, destination IP address, destination port number and the protocol in each flow traffic. Incorporating with supervised learning method they used, Nearest Neighbor classifier, several bags instead of numerous traffic flows are presented as input training data set. Figure \ref{fig:f4correlation} presents that the correlation technique contributes to their prediction phase, by combining the predicted results that the trained classifier get and the flow correlation analysis with vote mechanism. That is, the majority of predicted results for a set of network traffic data in one same bag, will be the final classification results of all those data. A probabilistic framework is built for BoF model to reduce the average classification error in each bag's prediction according to the Bayesian decision theory.

In terms of the challenge of correlation technique, we can conclude a few points. It still remains an open question about the way of identifying the new correlation based on the knowledge of traffic domain. The second problem is on how to correlate information among traffic data collected from different sources \cite{dainotti2012analysis}, which the data fusion problem should be further investigated. 

\subsection{Comparison of Recent Work}
As Table~\ref{tab:compare} illustrated, among these five key works mentioned above, four applied flow statistical-based traffic classification. Most of works including Zhang et al.~\cite{zhang2015robust} and Wang et al.~\cite{wang2014internet} employed semi-supervised learning among traffic flow classification to obtain the satisfactory performance. In addition, the popular correlated information that they found is 3-tuple heuristics: destination IP, destination port, and protocol. According to Ma et al.~\cite{ma2006unexpected}, packet contents can also be considered as correlation based on payload-based traffic classification. And no matter which the correlation technique is combined with supervised learning algorithms \cite{zhang2015robust} or unsupervised learning algorithms \cite{wang2014internet}, they all improved accuracy rate for traffic classification comparing with using machining learning algorithm alone. Additionally, to solve the unknown class problem, the robust traffic classification should combine with unsupervised learning and supervised learning with as little manual intervention as possible. Wherein the unsupervised learning can cluster flows of the same class to obtain known and unknown classes, and supervised learning can classify flows with unknown classes labeled manually. It would be ideal to further reduce the level and the frequency and intensity of the manual intervention.

\section{Discussion on Future work}
Although the state-of-art technologies attain cyber security to some extend, there are challenges. Such challenges will require more attention with respect to cyber security problems. 

As for the robutness technique, RTC mechanism assumes that the labeling results that human expert is of high accuracy, which is problematic due to the availability of the experts and the unavoidable human errors. Then classifiers are suggested to be retrained for a long period of time. Therefore, the system updated is triggered not only when new classes should be equipped  but also when previous known classes are learned again with new set of data training. And the costs in both time and money will increase when labors are involved with examining many unknown classes. To enhance the level of efficiency of RTC, sub-bag of flow is proposed by Zhang et al.~\cite{zhang2015robust}. Specifically, instead of using only one bag consisting of correlated flows, several sub-bags are constructed inside of one bag. And the flows in each sub-bag are collected from the same user and are sharing with 4-tuple rather than 3-tuple in BoF by adding an extra protocol-related feature. 

In addition, the correlations among flows remain mostly underused, though some of correlated information had already identified, like 5-tuple feature among traffic flow including source IP, destination IP, source port, destination port and protocol \cite{zhang2013network}. For example, the attacker's behavior that is analyzed and reflected on traffic flow data can be generalized and classified as a key feature in detecting malicious traffic. And this key feature can also be regarded as a piece of correlated information.

The unbalanced data source may affect the accuracy of the classification results. For instance, the quantity of class BitTorrent among one collected training data set is over 10,000 which is 1,000 times larger than class RTSP. Hence, many classifiers can easily extract the characteristics of class BitTorrent with strong support. Conversely, the 10 RTSP flows are too few to provide sufficient information for classifiers. As a result, the accuracy in classification of class RTSP will be far lower than the accuracy of class BitTorrent performed by the same classifier. Moreover, since the malicious and offensive applications almost always maintain minority, the big data with unbalanced classes structure makes the traffic classification methods difficult to identify abnormal traffic with a small amount among massive network traffic. To significantly improve the performance of classifier, we may have to introduce a preprocessing step when the portion of each class can be controlled manually before finalizing training data set.

\section{Conclusion}
In this paper, we started with introducing traffic analytics for cyber security applications from three aspects --- intrusion detection, malware analysis, and botnet detection.  In general, all detection applications employ various traffic classification to distinguish the abnormal traffic from other normal traffics identified from prior knowledge.  To enhance the cyber security, some techniques are summarized with detailed description.  Different from two packet-based classifications, flow statistical-based traffic classification is a new important tool for current big data analysis.  Flow statistical-based traffic classification not only can tackle the problem of various and changeable packets, but also is able to handle the encrypted traffic flow which is popular in use of contemporary and complex network.  Robust network traffic classification can detect unknown traffic, which is an essential improvement in cyber security for most malicious traffic is unknown from current systems. Such detection requires both unsupervised learning and supervised learning. Thus, the performance of classifiers can be significantly enhanced by using correlated information derived from various methods and integrated with machine learning algorithms. Furthermore, we discussed the limitations of state of the art techniques and pointed out future work. All in all, continuous research efforts are needed to improve automatic traffic analytics and to apply it broadly to address critical cyber security issues.



\ifCLASSOPTIONcaptionsoff
  \newpage
\fi



\bibliographystyle{IEEEtran}
\bibliography{library}

\end{document}